\documentclass[twocolumn,draft,prd,
showpacs,floatfix,preprintnumbers,nofootinbib]{revtex4}

\usepackage[final]{epsfig}
\usepackage{epsfig,graphicx}
\usepackage{graphicx}
\usepackage{bm}
\usepackage[usenames]{color}
\usepackage{lipsum}

\begin{document}

\title{Inspecting the supernova--gamma-ray burst connection with high-energy neutrinos}

\author{Irene Tamborra\footnote{Present address: Niels Bohr International Academy, Niels Bohr Institute, Blegdamsvej 17, 2100 Copenhagen, Denmark}}
\affiliation{GRAPPA Institute, University of Amsterdam, 1098 XH, Amsterdam, The Netherlands}

\author{Shin'ichiro Ando}
\affiliation{GRAPPA Institute, University of Amsterdam, 1098 XH, Amsterdam, The Netherlands}

\date{\today}

%%%%%%%%%%%%%%%%%%%%%%%%%%%%%%%%%%%%%%%%%%%%%%%%%%%%%%%%%%%%%%%%%%%%%%
\begin{abstract}
Long-duration gamma-ray bursts (GRBs) have been often considered as the
 natural evolution of some core-collapse supernovae (SNe).
While GRBs with relativistic jets emit an electromagnetic signal, GRBs with mildly relativistic jets are opaque to
 photons  and, therefore, could be detectable through neutrinos only.
We discuss the possibility that successful GRBs and mildly relativistic jets belong to the
 same class of astrophysical transients with different Lorentz factor
 $\Gamma_b$ and study the production of high-energy neutrinos as a
 function of $\Gamma_b$, by including both proton-photon and
 proton-proton interactions.
By assuming a SN--GRB connection, we find that the diffuse neutrino
 emission from optically thick jets with  Lorentz factors  lower than the ones of  successful GRBs
 can be one of the main components of the observed IceCube high-energy
 neutrino flux.
 Moreover, under the assumption that 
   all these jets  belong to the same class of astrophysical transients, we  show that the IceCube high-energy neutrino data provide
 indirect constraints on the rate of  non-successful jets, favoring a local  rate  lower than tens of
percent of the local SN rate.
These limits are currently comparable to the ones obtained in dedicated searches on choked
 sources and are expected to become tighter with accumulation of more
 high-energy neutrino data.
\end{abstract}
%%%%%%%%%%%%%%%%%%%%%%%%%%%%%%%%%%%%%%%%%%%%%%%%%%%%%%%%%%%%%%%%%%%%%%

%14.60.Lm	Ordinary neutrinos
%95.30.Cq	Elementary particle processes
%95.85.Ry	Neutrino, muon, pion, and other elementary particles; cosmic rays
%98.70.Rz	gamma-ray sources; gamma-ray bursts

\maketitle

%%%%%%%%%%%%%%%%%%%%%%%%%%%%%%%%%%%%%%%%%%%%%%%%%%%%%%%%%%%%%%%%%%%%%%
\section{Introduction}
%%%%%%%%%%%%%%%%%%%%%%%%%%%%%%%%%%%%%%%%%%%%%%%%%%%%%%%%%%%%%%%%%%%%%%
Gamma-ray bursts (GRBs) are among the most energetic astrophysical transients~\cite{Meszaros:2015krr,Kumar:2014upa,Meszaros:2006rc} 
and possibly sources of ultra-high-energy cosmic rays~\cite{Waxman:1995vg,Vietri:1995hs}.
Several observations point toward a connection between long-duration GRBs and core-collapse supernovae (SNe)~\cite{Woosley:2006fn}. 
The most likely scenario is that an accretion disk surrounding a black
hole forms soon after the core collapse and a jet
originates~\cite{Woosley:1993wj}.
Such hypothesis is supported from the fact that SNe and GRBs are
expected to release a comparable amount of kinetic energy.

Once the jet is formed, two extreme scenarios may be foreseen: Either
the jet is highly or mildly relativistic.
The former corresponds to an ordinary long-duration GRB with Lorentz
factor $\Gamma_b$ of $\mathcal{O}(10^2$--$10^3)$, where the emission of
high-energy neutrinos is accompanied by an electromagnetic counterpart.
The latter stands for a baryon-rich jet with Lorentz factor of a few,
where no electromagnetic counterpart is expected, the jet is optically
thick, and only neutrinos are able to escape;  in the following, these jets will be dubbed 
as  ``baryon-rich or low-$\Gamma_b$ jets.''

The IceCube telescope at the South Pole should be sensitive to any of
the above classes of GRBs, given their sizeable neutrino production.
However, all the GRB dedicated searches performed from IceCube over the
past years reported negative
results~\cite{Abbasi:2009ig,Abbasi:2011qc,Abbasi:2012zw,Aartsen:2014aqy,Abbasi:2011ja,Aartsen:2015pwd},
constraining the  theoretical models employed to explain
the GRB neutrino emission~\cite{Baerwald:2011ee,He:2012tq,Bartos:2013hf}.

The IceCube experiment recently detected astrophysical neutrinos with the highest
energies ever observed~\cite{Aartsen:2013bka,Aartsen:2013jdh,
Aartsen:2014gkd,Aartsen:2014muf, Aartsen:2015knd, Aartsen:2015rwa}.
The current data set is compatible with an equal distribution of such
neutrinos in flavor and an isotropic allocation in the sky.
Various astrophysical sources such as starburst galaxies, low-luminosity
gamma-ray bursts, and active galactic nuclei have been considered as they might produce a neutrino
flux comparable to the detected
one~\cite{Meszaros:2015krr,Waxman:2015ues,Murase:2015ndr,Anchordoqui:2013dnh,Tamborra:2014xia}.
However, more recent analysis point toward a lower neutrino flux from
star-forming galaxies and blazars than previously 
expected, see e.g. Refs.~\cite{Ando:2015bva,Bechtol:2015uqb,Glusenkamp:2015jca}.
Such recent developments leave open the quest on the origin of the
IceCube high-energy events, and suggest faint or low-luminosity sources as plausible components of the
measured flux (see also discussion in Refs.~\cite{Murase:2015xka,Kistler:2015ywn}), besides yet unknown astrophysical sources.
The diffuse neutrino emission from low-luminosity
GRBs~\cite{Waxman:1998yy,Murase:2006mm,Gupta:2006jm,Liu:2011cua} could
indeed partly explain the IceCube PeV
events~\cite{Liu:2012pf,Murase:2013ffa,Tamborra:2015qza,Fraija:2015nsa}.

In this paper, we assume that high- and low-$\Gamma_b$ jets belong to
the same
GRB family~\cite{Nakar:2015tma,Bartos:2012sg,Guetta:2006gq,Woosley:2006fn,MacFadyen:1998vz,Paczynski:1997yg,Woosley:1993wj}
and that the local rate of such jets progressively increases as
$\Gamma_b$ decreases, reflecting the fact that the  baryon-rich sources could be
more abundant than the  ordinary GRBs~\cite{Meszaros:2001ms,Razzaque:2003uv}.
We expect that PeV neutrinos are mostly produced through proton-photon
($p\gamma$) interactions in jets with high $\Gamma_b$, while TeV
neutrinos are emitted from baryon-rich sources mostly by means of
proton-proton ($pp$)
interactions~\cite{Meszaros:2001ms,Razzaque:2003uv,Razzaque:2005bh,Ando:2005xi,Horiuchi:2007xi}.
In order to estimate the high-energy neutrino emission from
astrophysical jets as a function of $\Gamma_b$, we model the neutrino
emission as generally as possible and by including both $pp$ and
$p\gamma$ interactions.
We tune the local rate of high-$\Gamma_b$ bursts to the observed one of
high-luminosity GRBs and assume that the one of the low-$\Gamma_b$ jets
is a fraction of the local core-collapse SN rate.

The purpose of our work is to investigate whether IceCube high-energy neutrino data can
constrain the SN--GRB connection and, possibly, can allow us to extrapolate upper bounds
on the abundance of baryon-rich jets.
We also discuss the range where $pp$ and $p\gamma$ interactions 
dominate the neutrino production as a function of the Lorentz factor $\Gamma_b$ and find that
optically thick jets with  $\Gamma_b$  lower than the one of successful GRBs, e.g.~$\Gamma_b \sim \mathcal{O}(10-100)$, could provide a
substantial contribution to the observed IceCube flux of high-energy
neutrinos.

The paper is organized as follows.
In Sec.~\ref{sec:astrojetmodel}, we define the jet emission model and
estimate the expected high-energy neutrino production from these
sources.
In Sec~\ref{sec:diffuse}, we study the diffuse emission of high-energy
neutrinos from astrophysical jets as a function of $\Gamma_b$. 
Bounds on the physics of baryon-rich jets through the IceCube
high-energy neutrino data are discussed in Section~\ref{sec:radiation} as well as the
dependence of our results from the jet model parameters.
Outlook and conclusions are presented in Sec.~\ref{sec:conclusions}.  

%%%%%%%%%%%%%%%%%%%%%%%%%%%%%%%%%%%%%%%%%%%%%%%%%%%%%%%%%%%%%%%%%%%%%%
\section{High-energy neutrino production in relativistic jets}
\label{sec:astrojetmodel}
%%%%%%%%%%%%%%%%%%%%%%%%%%%%%%%%%%%%%%%%%%%%%%%%%%%%%%%%%%%%%%%%%%%%%%
In this Section, we define a generic model for the high-energy neutrino
emission from astrophysical relativistic jets by including $pp$ and
$p\gamma$ interactions.
Besides studying the cooling of protons as a function of $\Gamma_b$, we
discuss cooling processes of pions, kaons and muons as well as the
expected neutrino fluence.

%%%%%%%%%%%%%%%%%%%%%%%%%%%%%%%%%%%%%%%%%%%%%%%%%%%%%%%%%%%%%%%%%%%%%%
\subsection{Jet emission properties}                \label{sec:jetmodel}
%%%%%%%%%%%%%%%%%%%%%%%%%%%%%%%%%%%%%%%%%%%%%%%%%%%%%%%%%%%%%%%%%%%%%%
Independently from the bulk Lorenz factor $\Gamma_b$, we consider a
typical jet with total energy $\tilde{E}_j \sim 3 \times
10^{51}$~erg~\footnote{For each
 quantity $X$, we adopt $\tilde{X}$, $X^\prime$ and $X$ for the
physical quantity defined in the source frame, in the jet comoving frame
and in the observer frame respectively.}, jet opening angle $\theta_j \sim 5$ degrees, total
duration $\tilde{t}_j \sim 10$~s, and  total jet luminosity $\tilde{L}_j
= \tilde{E}_j/\tilde{t}_j$~\cite{Meszaros:2006rc}.
Internal shocks between the jet plasma ejecta occur at 
$\tilde{r}_j \sim 2 \Gamma_b^2 c t_v/(1+z)$ with  $t_v \sim 0.1$~s the
jet variability time, $c$ the speed of light and $z$ the redshift.
The isotropic luminosity carried by photons in successful bursts is
$\tilde{L}_{\rm iso} = 2 (1+z)^2 L_j \epsilon_e/(0.3
\theta_j^2)$~~\footnote{
As pointed out in Ref.~\cite{Kakuwa:2011aq}, the scaling relation between $\tilde{L}_{\rm iso}$ and $L_j$ is characterized by a large 
uncertainty that we assume to be fixed to its best fit value ($0.3\pm0.2$); variations within the allowed uncertainty band may
be responsible for changes in the typical energy of non-thermal photons.} with $\epsilon_e=0.1$ the energy
fraction carried by electrons~\cite{Meszaros:2006rc,Tamborra:2015qza,Kakuwa:2011aq}.
In general, one could expect the jet properties to vary as a function of
$\Gamma_b$.
However, we do not have data on baryon-rich jets and assume that their properties are
on average comparable to the ones of successful bursts for the sake of
simplicity. 
The similarity between kinetic energies of relativistic GRB jets and
non-relativistic SN explosions may support such an assumption. 
Nevertheless, we refer the interested reader to Sec.~\ref{sec:radiation} for a discussion
on the dependence of our results on the adopted model parameters.

In order to characterize the typical photon energy, we introduce the
Thomson optical depth~\cite{Razzaque:2005bh}:
\begin{equation}
\tau_T^\prime = \frac{\sigma_T n^{\prime}_e \tilde{r}_j}{\Gamma_b}\ ,
\label{eq:tauT}
\end{equation}
with the comoving electron  density similar to the one of baryons
\begin{equation}
n^{\prime}_{e} \simeq n^{\prime}_{p}   = \frac{L_j (1+z)^4}{m_p c^5
 \Gamma_b^6 8 \pi \theta_j^2 t_v^2}\ ,
\end{equation}
where $n^{\prime}_{p}  = [E_j (1+z)]/(m_p c^2 \Gamma_b V^\prime)$, $m_p$ is the proton mass, and
$V^\prime = 2 \pi \theta^2_j \Gamma_b \tilde{r}^2_j c t_j/(1+z)$ is the comoving volume.

The energy associated to the jet magnetic field $B$ is
\begin{equation}
\frac{B^{\prime 2}}{8 \pi} \simeq \frac{2 \epsilon_B E^\prime_j (1+z)}{\pi \theta_j^2 \tilde{r}^2_j c t_j \Gamma_b}\ ,
\end{equation} 
$\epsilon_B \simeq 0.1$ being the fraction of the total energy converted into
 magnetic energy. Note that we adopt $B^{\prime 2}/(8 \pi)=4 \epsilon_B E^\prime_j/V^\prime$~\cite{Tamborra:2015qza,Meszaros:2006rc}. Previous work on the topic does not always include the constant numerical factor in the definition of the magnetic energy density (see, e.g., Ref.~\cite{Baerwald:2011ee}); therefore care should be taken when comparing our results with the ones reported in part of the existing literature.

In the case of optically thin sources ($\tau_{T}^\prime < 1$),
the photon energy distribution is non-thermal with a typical energy 
$E^\prime_\gamma=(\hbar \epsilon_e^2 m_p^2 e B^\prime)/(m_e^3 c)$~\cite{Waxman:2003vh}, i.e.,
\begin{equation}
E^\prime_{\gamma, {\rm non-th}} \simeq \frac{(1+z)  \epsilon_{e,-1}^{3/2}  \epsilon_{B,-1}^{1/2} \tilde{L}_{{\rm iso},52}^{1/6}}{\Gamma_b t_{v,-2}^{2/3}}~\mathrm{MeV}\ ,
\end{equation}
with $\epsilon_{e} \simeq 0.1$ the energy  fraction carried by the electrons. 
The subscripts in the above equation stand for the typical order of
magnitude of the quantities defining $E^\prime_{\gamma, {\rm non-th}}$,
i.e. $\epsilon_{e,-1}=\epsilon_{e}/0.1$, $\tilde{L}_{{\rm
iso},52}=\tilde{L}_{{\rm iso}}/(10^{52}\ \mathrm{erg})$ and similarly
for the other variables.

 On the other hand, when $\tau_{T}^\prime \ge 1$, photons thermalize and have an average black-body temperature~\cite{Razzaque:2005bh}:
\begin{equation}
E^\prime_{\gamma, {\rm th}} \simeq \left(\frac{15 \hbar^3 c^2 \epsilon_e E_j}{2 \pi^4 \tilde{r}_j^2  t_j}\right)^{1/4}\ .
\end{equation}
  Therefore, the most general expression for the characteristic photon energy is 
\begin{equation}
\label{eq:photonbreak}
E^\prime_\gamma = \left\{ \begin{array}{lll}
E^\prime_{\gamma, {\rm th}}  
&\mathrm{for} & \tau_T^\prime > 1 \\ 
E^\prime_{\gamma, {\rm non-th}}  & \mathrm{for} &  \tau_T^\prime \le 1\ .
\end{array}\right.
\end{equation}
Note that, although Eq.~(\ref{eq:photonbreak}) defines characteristic photon energies, 
we include specific spectral shapes of the non-thermal and thermal 
photon energy spectrum in the numerical computations as well as in the following
discussion (including Fig.~\ref{fig:nuinitial}).

%%%%%%%%%%%%%%%%%%%%%%%%%%%%%%%%%%%%%%%%%%%%%%%%%%%%%%%%%%%%%%%%%%%%%%
\subsection{Proton acceleration and cooling processes}                \label{sec:proton}
%%%%%%%%%%%%%%%%%%%%%%%%%%%%%%%%%%%%%%%%%%%%%%%%%%%%%%%%%%%%%%%%%%%%%%

The acceleration time (acc) of a proton with comoving energy $E_p^\prime$ is:
\begin{equation}
t^\prime_{\rm acc} = \frac{E^\prime_p}{B^\prime e c} \simeq 3.7 \times 10^{-9} {\mathrm s} \frac{E^\prime_{p,\mathrm{GeV}} \theta_{j,7} t_{v,-2} \Gamma_{b,2.5}^3 t_{j,1}^{1/2}}{\epsilon_{B,-1}^{1/2} E_{j,51}^{1/2} (1+z)^2}\ ,
\end{equation}
 under the assumption of perfectly efficient acceleration.
In the presence of a magnetic field, protons are subject to synchrotron cooling (sync) besides being accelerated: 
\begin{equation}
\label{eq:synccooling}
t^\prime_{\rm sync}=\frac{3 m_p^4 c^3 8 \pi}{4 \sigma_T m_e^2 E^\prime_p B^{\prime 2}}\simeq5 \times 10^{9} {\mathrm s} \frac{\theta_{j,7}^2 t_{j,1}\Gamma_{b,2.5}^6 t_{v,-2} E_{j,51}^{-1}}{E^\prime_{p,\mathrm{GeV}} \epsilon_{B,-1}  (1+z)^4}.
\end{equation}

The Inverse Compton process (IC) is another cooling channel. We express it as a function of $E^\prime_p$ for the Thomson and Klein-Nishina
regimes~\cite{Razzaque:2005bh}:
\begin{equation}
\label{eq:ICcooling}
t^\prime_{\rm IC}=\left\{\begin{array}{lll} \frac{3 m_p^4 c^3}{4 \sigma_T m_e^2 E^\prime_p E^\prime_\gamma n^\prime_\gamma} \simeq 5\times10^9 {\mathrm s} \frac{\theta_{j,7}^2 t_{j,1}\Gamma_{b}^6 t_{v,-2}^2 E_{j,51}^{-1}}{E^\prime_{p,GeV}  \epsilon_{e,-1} (1+z)^4}\ ,\\
\frac{3 E_p^\prime E^\prime_\gamma}{4 \sigma_T m_e^2 c^5
 n^\prime_\gamma}\simeq 6\times10^3{\mathrm
 s}\frac{E_{p,\mathrm{GeV}}^\prime E_{\gamma \mathrm{MeV}}^{\prime 2}
 \theta^2_{j, 7} \Gamma_{b,2.5}^6 t_{v,-2}^2}{E_{j,51} \epsilon_{e,-1}
 (1+z)^4 t_{j,-1}^{-1}}\ ,  
\end{array}\right.
\end{equation}
respectively for $E^\prime_p \ll (\gg) m_p^2 c^4/E^\prime_\gamma$, and  with the comoving photon density 
\begin{equation}
n^{\prime}_{\gamma} = \frac{2 E^\prime_j \epsilon_e (1+z)}{\pi \theta^2_j \tilde{r}^2_j c t_j \Gamma_b E^\prime_\gamma}\ .
\end{equation}

Because of the high density of photons, $e^+e^-$ pairs may be produced through the  
Bethe-Heitler (BH) process: $p\gamma \rightarrow p e^+ e^-$. The BH cross-section is defined  as in, e.g., Ref.~\cite{Razzaque:2004yv}:
\begin{equation}
\sigma_{\rm BH}=\alpha r_e^2 \left\{\frac{28}{9} \ln\left[\frac{2 E^\prime_p E^\prime_\gamma}{m_p m_e c^4}\right]-\frac{106}{9}\right\}\ ,
\end{equation}
with $\alpha$ the fine-structure constant and $r_e$ the classical electron radius; the corresponding comoving cooling time is~\cite{Razzaque:2005bh}
\begin{equation}
t^\prime_{\rm BH} = \frac{E^\prime_p (m_p^2 c^4 + 2 E^\prime_p E^\prime_\gamma)^{1/2}}{2 n^\prime_\gamma \sigma_{\rm BH} m_e c^3 (E^\prime_p+E^\prime_\gamma)}\ .
\end{equation}

The characteristic times of $p\gamma$ and $pp$ interactions can be of relevance for the proton cooling, besides producing high-energy neutrinos
as we will see in the next section. These are: 
\begin{equation}
t^\prime_{p\gamma} = \frac{E^\prime_p}{c \sigma_{p\gamma} n^\prime_\gamma \Delta E^\prime_p} \simeq 1.5 \times 10^4 \mathrm{s} \frac{\theta_{j,7}^2 \Gamma_{b,2.5}^6 t_{v,-2}^2 t_{j,1} E^\prime_{\gamma,\mathrm{MeV}}}{E_{j,51} \epsilon_{e,-1} (1+z)^4}\ , 
\end{equation}
\begin{equation}
t^\prime_{pp} = \frac{E^\prime_p}{c \sigma_{pp} n^\prime_p \Delta E^\prime_p}\simeq 1.4 \times 10^4 \mathrm{s} \frac{\Gamma_{b,2.5}^6 \theta^2_{j,7} t^2_{v,-2} t_{j,1}}{E_{j,51} (1+z)^4}\ ,
\end{equation}
with cross-sections $\sigma_{p\gamma} \simeq 5 \times 10^{-28}$~cm$^{-2}$ and $\sigma_{pp} \simeq 5 \times 10^{-26}$~cm$^{-2}$ 
respectively, and $\Delta E_p^\prime/E^\prime_p = 0.2\ (0.8)$ for $p\gamma$ ($pp$)
interactions~\cite{Razzaque:2005bh}.  In the above equations, we neglect the integral over the energy range of the 
cross sections as well as the multi-pion production for sake of simplicity. Such approximations are not crucial given the purpose of our study, but we refer the interested reader to  Refs.~\cite{Murase:2006dr,Baerwald:2010fk,Baerwald:2011ee} for a discussion on the role of these factors on the total expected neutrino flux. Note that the expression for
$t^\prime_{p\gamma}$ is valid for $E^\prime_p$ above the 
 threshold  energy of the $\Delta$ resonance due to $p\gamma$ interactions ($E^\prime_{p,{\rm th}} = [(m_\Delta c^2)^2-(m_p c^2)^2]/4 E^\prime_\gamma$)~\cite{Guetta:2003wi,Tamborra:2015qza}.

Finally, protons are subject to the adiabatic cooling (ac), whose time scale
in the comoving frame is given by
\begin{equation}
t^\prime_{\rm ac}=\frac{\tilde{r}_j}{\Gamma_b c}\simeq 6.3\ \mathrm{s}\
 \Gamma_{b,2.5} t_{v,-2} (1+z)^{-1}\ .
 \label{eq:AC}
\end{equation}

The total cooling time of protons in the comoving frame is given by the superposition of the cooling processes mentioned above:
\begin{equation}
t^{\prime\ -1}_{c,p} = t^{\prime\ -1}_{\rm syn}+t^{\prime\ -1}_{\rm
 IC}+t^{\prime\ -1}_{\rm BH}+t^{\prime\ -1}_{p\gamma}+t^{\prime\
 -1}_{pp}  + t^{\prime\ -1}_{\rm ac} \ .
\end{equation}
The maximum proton energy ($E_{p,{\rm max}}$) is determined by
\begin{equation}
t^{\prime}_{c,p}=t^\prime_{\rm acc}\ .
\end{equation} 
While the minimum proton energy ($E_{p,{\rm min}}$) is given by the proton rest mass. 

\begin{figure}
\includegraphics[width=1.0\linewidth]{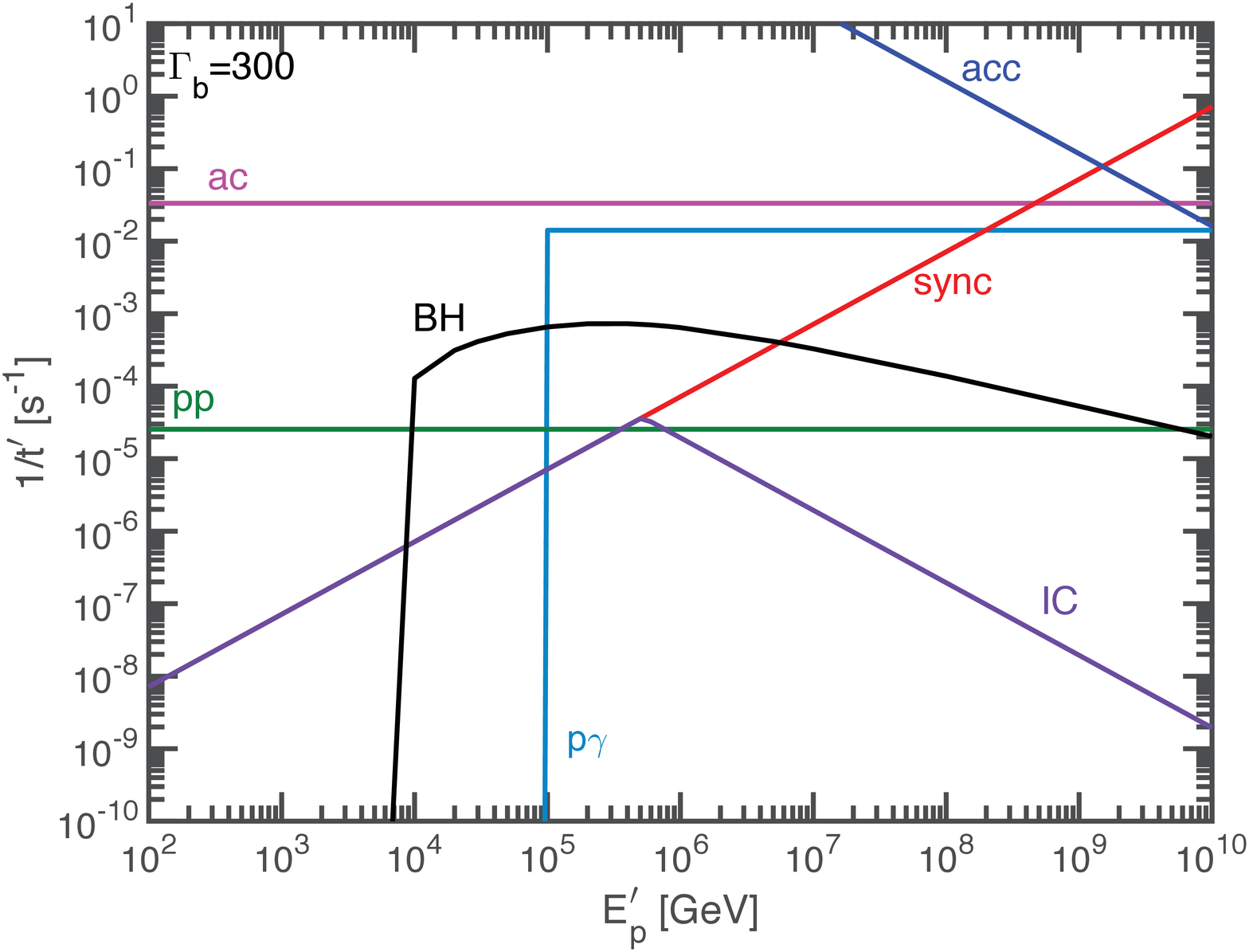}\\
\includegraphics[width=1.\linewidth]{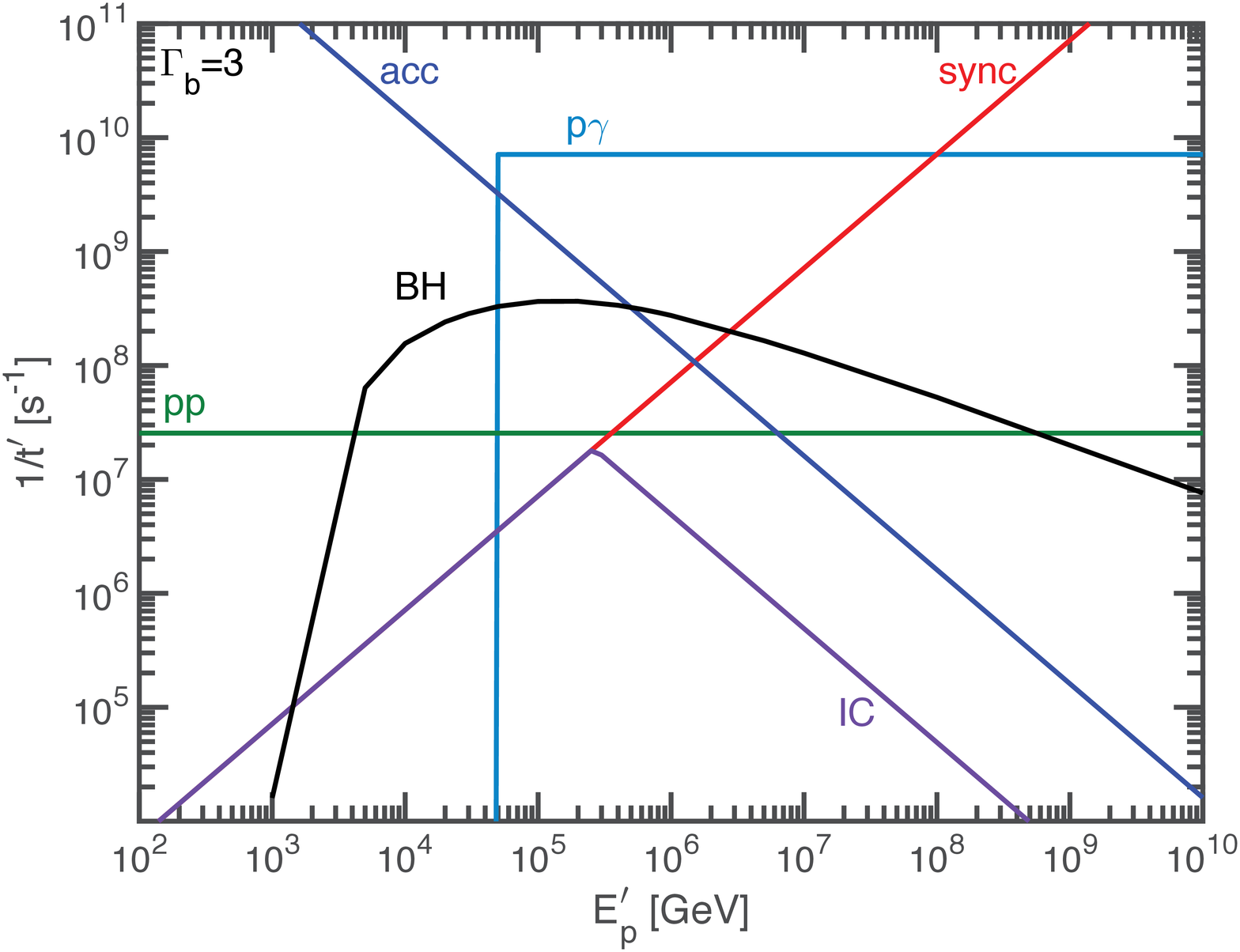}
\caption{\label{fig:protoncooling} Comoving  cooling times for photomeson ($1/t^\prime_{p\gamma}$ in light blue), proton-proton ($1/t^\prime_{pp}$ in green), 
synchrotron radiation ($1/t^\prime_{\rm syn}$ in red), inverse Compton ($1/t^\prime_{\rm IC}$ in violet), BH ($1/t^\prime_{\rm BH}$ in black), and adiabatic 
 cooling ($1/t^\prime_{\rm ac}$ in magenta)  as functions of the comoving proton energy $E^\prime_p$. The shock acceleration time ($1/t^\prime_{\rm acc}$) is plotted in 
 blue for comparison. The top (bottom) panel shows a typical jet with $\Gamma_b=300$ ($\Gamma_b=3$) at 
$z=1$. The maximum proton energy is determined by $t^\prime_{\rm syn}$ for $\Gamma_b=300$ and by $t^\prime_{p\gamma}$ for $\Gamma_b=3$.
Note as the adiabatic cooling time is below the minimum y-axis value  in the $\Gamma_b=3$ plot.} 
\end{figure}

In order to provide an idea of the relevant cooling processes for
protons for both high and low $\Gamma_b$'s, Fig.~\ref{fig:protoncooling}
shows (inverse of) the proton cooling time scales  as functions 
of the comoving proton energy for $\Gamma_b=300$ (top panel) and
$\Gamma_b=3$ (bottom) jets at $z=1$.
In the high-$\Gamma_b$ case, the maximum proton energy is 
$E^\prime_{p,{\rm max}}=1.5 \times 10^9$~GeV and it is determined by the
competition of the shock acceleration with synchrotron cooling
characterized by $t^\prime_{\rm sync}$. The adiabatic cooling becomes instead the 
dominant cooling process for $\Gamma_b$  higher than the one shown here.  The low-$\Gamma_b$ case has a 
maximum proton energy $E^\prime_{p,{\rm max}}=9.5 \times 10^4$~GeV fixed
by the photomeson cooling  ($t^\prime_{p\gamma}$). 
Note that $pp$ interactions occur for the whole proton energy range
independently from $\Gamma_b$, while $p\gamma$ interactions are relevant
for $E^\prime_p$ larger than the threshold energy of the $\Delta$
resonance.
However, as discussed in the next section, the energy range where
$p\gamma$ interactions are relevant becomes smaller as $\Gamma_b$
decreases, until $p\gamma$ interactions are negligible for low
$\Gamma_b$ and only $pp$ interactions affect the neutrino spectrum (see
case shown in the bottom panel of Fig.~\ref{fig:protoncooling}).

%%%%%%%%%%%%%%%%%%%%%%%%%%%%%%%%%%%%%%%%%%%%%%%%%%%%%%%%%%%%%%%%%%%%%%
\subsection{Meson cooling processes and neutrino production}                \label{sec:meson}
%%%%%%%%%%%%%%%%%%%%%%%%%%%%%%%%%%%%%%%%%%%%%%%%%%%%%%%%%%%%%%%%%%%%%%
High energy neutrinos are produced by the  protons interacting with the 
synchrotron photons ($p\gamma$ interactions) and with the protons present in
the shock region ($pp$
interactions)~\cite{Kelner:2006tc,Kelner:2008ke}.
Both interactions  produce $\pi^{\pm}$ and $K^{\pm}$ that then decay to
muons and neutrinos.

Pions and kaons are affected by hadronic cooling:
\begin{equation}
t^\prime_{\rm hc} = \frac{E^\prime_a}{\Delta E^\prime_a c \sigma_{h} n_p^\prime}= t^\prime_{\rm pp}\ .
\end{equation}
with $\Delta E^\prime_\pi = 0.8 E^\prime_\pi$ the energy lost by the
incident meson in each collision (Figs.~18 and 19 in Ref.~\cite{Brenner:1981kf}), and $\sigma_{h}= 5 \times 10^{-26}$~cm$^{-2}$ the cross-section for meson-proton collisions~\cite{Eidelman:2004wy}.
Similarly to protons, pions, kaons and muons are also subject to
synchrotron and IC cooling (defined as in
Eqs.~\ref{eq:synccooling} and \ref{eq:ICcooling} with $E^\prime_p
\rightarrow E^\prime_a$ and $m_p \rightarrow m_a$, where $a=\pi$, $K$,
and $\mu$) as well as to adiabatic cooling
(Eq.~\ref{eq:AC}).

The total cooling time of pions, kaons and muons is:
\begin{equation}
t^{\prime\ -1}_c = t^{\prime\ -1}_{\rm syn}+ t^{\prime\ -1}_{\rm hc}+t^{\prime\ -1}_{\rm IC}+t^{\prime\ -1}_{\rm ac}\ .
\end{equation}
(But note that muons are not subject to hadronic cooling.)
As explained in the next section, by comparing the above cooling times
with the lifetime of mesons and muons, one can predict the expected
neutrino spectrum.

%%%%%%%%%%%%%%%%%%%%%%%%%%%%%%%%%%%%%%%%%%%%%%%%%%%%%%%%%%%%%%%%%%%%%%
\subsection{Neutrino energy spectrum}                \label{sec:neutrinoenergyspectrum}
%%%%%%%%%%%%%%%%%%%%%%%%%%%%%%%%%%%%%%%%%%%%%%%%%%%%%%%%%%%%%%%%%%%%%%

\begin{figure}
\includegraphics[width=.8\linewidth]{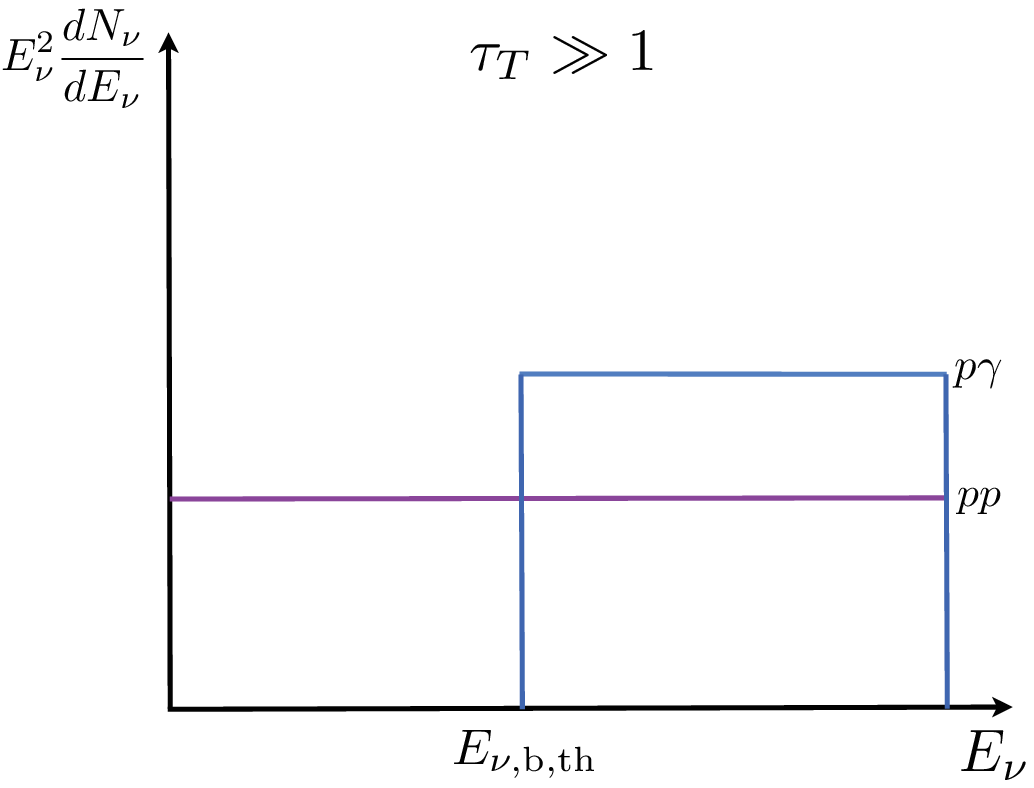}\\
\includegraphics[width=.8\linewidth]{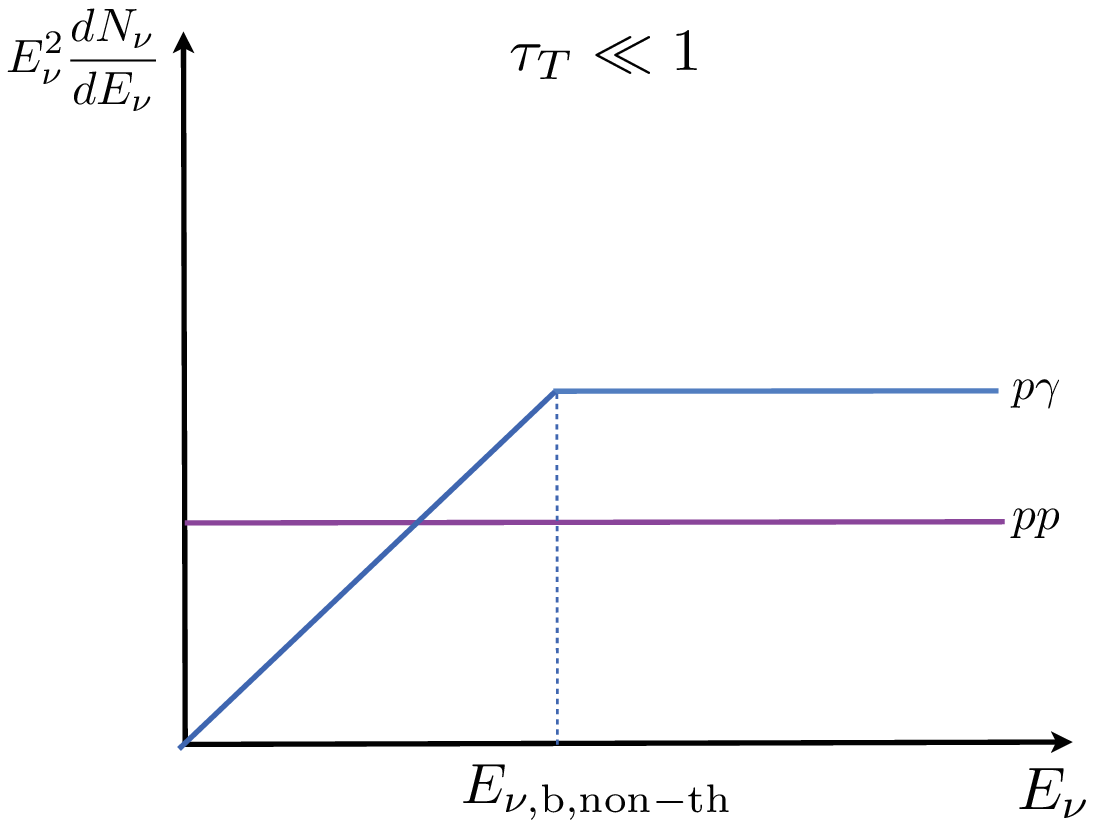}
\caption{\label{fig:nuinitial}Diagram for the neutrino energy spectrum in absence of meson cooling effects. 
The top (bottom) panel  refers to the $\tau_T \gg 1$ ($\tau_T \ll 1$) case. For each case, the spectra from
$pp$ and $p\gamma$ interactions are shown in arbitrary units. The relative normalization of the $pp$ spectrum
with respect to the $p\gamma$ one is also arbitrary.}
\end{figure}

We assume an initial proton spectrum that scales as $E_p^{\prime -2}$.
As for $p\gamma$ interactions, the comoving proton energy to
produce a $\Delta$ resonance (and therefore to generate neutrinos) is
 $E^\prime_p \ge [(m_\Delta c^2)^2-(m_p c^2)^2]/(4
E^\prime_\gamma)$.
This energy will affect the correspondent neutrino spectrum at
\begin{eqnarray}
 E_{\nu, b} &=& a_i \left(\frac{\Gamma}{1+z}\right)^2
  \frac{(m_\Delta c^2)^2-(m_p c^2)^2}{4 E_{\gamma}}\nonumber\\
 &\simeq&  7.5 \times 10^5 \mathrm{GeV}\
  \left(\frac{a_i}{0.05}\right)
 \frac{\Gamma_{b,2.5}^2}{(1+z)^2 E_{\gamma,\mathrm{MeV}}}\ , 
\label{eq:nub1}
\end{eqnarray}
with $E_{\gamma}$ the characteristic energy of the photon spectrum (Eq.~6). The numerical factor $a_i$ is: $a_\pi = 0.05$ ($20\%$ being the
fraction of $E_p$ that goes into pions and $1/4$ is the
fraction of the $E_\pi$ carried by neutrinos), $a_{\mu_\pi} = 0.05$
(as $3/4$ is the energy fraction transferred from pions to muons and
$1/3$ takes into account the three-body decay of the muon), $a_K = 0.1$ ($20\%$
is  the fraction of $E_p$  that goes into $E_K$, and $1/2$
the fraction of $E_K$ carried by neutrinos), and  $a_{\mu_K}=0.033$
for muons originating from the kaon decay. 
Note as Eq.~(\ref{eq:nub1}) defines a spectral break in the neutrino energy spectrum for $\tau_T \ll 1$,
where the photon spectrum is a broken power law; it represents the threshold energy of the neutrino spectrum for the case 
with $\tau_T \gg 1$ (see Fig.~\ref{fig:nuinitial}).

The initial neutrino energy spectrum before being affected by meson cooling
processes is sketched in Fig.~\ref{fig:nuinitial} for $\tau_T \gg 1$
(top) and $\tau_T \ll 1$ (bottom).
For $\tau_T \gg 1$, given the sharp drop
of the black-body energy spectrum (see Sec.~\ref{sec:jetmodel}), the neutrino spectrum can be
approximated by a  rectangular function different from zero for $E^\prime_{\nu, b} < E^\prime_\nu <
E^\prime_{\nu, {\rm max}}$.
For $\tau_T \ll 1$, the neutrino spectrum will be the same as the proton
spectrum above $E^\prime_{\nu, b}$, while it will be harder at lower
energies (see also discussion in Refs.~\cite{Tamborra:2015qza,Guetta:2003wi}). 

Because of the cooling processes of pions, kaons, and muons described in
Sec.~\ref{sec:meson}, the neutrino spectrum is given by
\begin{equation}
\left(\frac{dN_\nu}{dE^\prime_\nu}\right)_{\rm inj} =
 \left(\frac{dN_\nu}{dE^\prime_\nu}\right)_0
 \left[1-\exp\left(-\frac{t^\prime_{c,a} m_a}{E^\prime_a
	      \tau_a}\right)\right]\ ,
 \label{eq:breaks}
\end{equation}
with $E^{\prime 2}_{\nu} (dN_\nu/dE^{\prime}_{\nu})_0$ defined as in
Fig.~\ref{fig:nuinitial} according to the value of $\tau_T$, and 
$\tau_a$ the lifetime of mesons or muons.
The minimum and maximum energies of the neutrino spectrum are defined by
the minimum and maximum proton energies introduced in
Sec.~\ref{sec:proton}.

The $\nu_e$ and $\nu_\mu$ neutrino energy spectra produced  from pion decay  are: 
\begin{equation}
\left(\frac{dN_{\nu_e}}{dE_\nu^\prime}\right)_{{\rm inj},\pi}=\left(\frac{dN_{\nu}}{dE_\nu^\prime}\right)_{\mu_\pi}\ ,
\end{equation}
\begin{equation}
\left(\frac{dN_{\nu_\mu}}{dE_\nu^\prime}\right)_{{\rm inj},\pi}=\left(\frac{dN_{\nu}}{dE_\nu^\prime}\right)_{\mu_\pi} + \left(\frac{dN_{\nu}}{dE_\nu^\prime}\right)_{\pi}\ .
\end{equation}
Similar relations hold for neutrinos produced by kaon decay. 
No $\nu_\tau$'s are  produced at the source. However, neglecting non-standard scenarios,   the three neutrino flavors become  similarly
abundant after flavor oscillations on their way to Earth~\cite{Anchordoqui:2013dnh}. 
Flavor conversions in matter might also
occur while neutrinos are propagating within the jet in optically thick
sources (see, e.g,
Refs.~\cite{Kashti:2005qa,Mena:2006eq,Fraija:2013cha,Xiao:2015gea,Fraija:2015gaa}). However, in the 
following, we will neglect flavor oscillations in matter as they would not affect
our conclusions. Nonetheless,
future constraints on the neutrino flavor ratio observed on Earth might provide us with
indirect information on the progenitor structure of optically thick
sources in the case of the successful observation of baryon-rich bursts.

By adapting the normalization proposed in Ref.~\cite{Hummer:2011ms},  the observed neutrino spectrum
for a single source at redshift $z$ and  for each neutrino production channel $a$ is:
\begin{eqnarray}
F_{\nu}(E_\nu, z) = \frac{(1+z)^3}{2 \pi \theta_j^2 \Gamma_b d^2_L}
 E^\prime_j N_a f_{p} [1-(1-\chi_p)^{\tau^\prime_p}]
 \left(\frac{dN_\nu}{dE^\prime_\nu}\right)_{{\rm inj}}\,
 \label{eq:fluence}
\end{eqnarray}
with $N_\pi=N_{\mu_\pi}=0.12$, $N_K=0.01$,
$N_{\mu_K}=0.003$~\cite{Tamborra:2015qza}, $\tau_p^\prime =
\tau_{pp}^\prime(z,E^\prime_\nu)+
\tau_{p\gamma}^\prime(z,E^\prime_\nu)= \tilde{r}_j/\Gamma_b (\sigma_{p\gamma} n^\prime_\gamma+\sigma_{pp} n^\prime_p)$~\cite{Razzaque:2005bh}, and $(dN_\nu/dE^\prime_\nu)_{\rm
inj}$ is the normalized spectrum that satisfies
$ \int dE^\prime_\nu E_\nu^\prime
  (dN_\nu/dE_\nu^\prime)_{\rm inj} = 1$.
The factor $f_p$ takes into account the effect of spectral breaks:
\begin{equation}
 f_{p}=
 \frac{\int_0^\infty dE^\prime_\nu E^\prime_\nu
 \left(\frac{dN_\nu}{dE^\prime_\nu}\right)_{{\rm
 inj}}}{\int_0^\infty dE^\prime_\nu E^\prime_\nu
 \left(\frac{dN_\nu}{dE^\prime_\nu}\right)_{{\rm
 inj,no-break}}}\ ,
\end{equation}
where $(dN_\nu/dE_\nu^\prime)_{\rm inj,no-break} \propto E_\nu^{\prime
-2}$ is the neutrino spectrum without any cooling breaks
(Eq.~\ref{eq:breaks}) as well as the break due to threshold of $p\gamma$
interaction (Eq.~\ref{eq:nub1}); the denominator is therefore
proportional to $\ln(E_{p,{\rm max}}^\prime/E_{p,{\rm min}}^\prime)$.
Finally, the neutrino energy in the jet frame is related to that in the
observer frame through $E^\prime_\nu=E_\nu (1+z)/\Gamma_b$, and
$d_L(z)$ is the luminosity distance defined in a flat $\Lambda$CDM
cosmology with $\Omega_m=0.32$, $\Omega_\Lambda=0.68$ and $H_0 =
67$~km~s$^{-1}$~Mpc$^{-1}$ for the Hubble constant~\cite{Ade:2013zuv}.

\begin{figure}
\includegraphics[width=1.\linewidth]{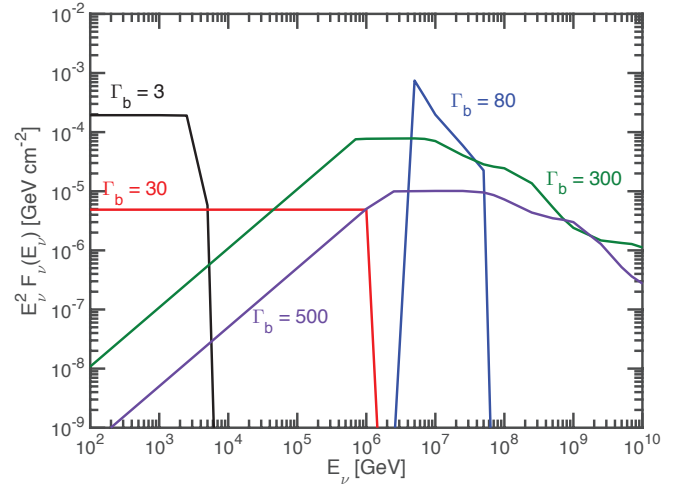}
\caption{\label{fig:singlesource} Expected fluence for a single flavor of an astrophysical burst at $z=1$ with $\Gamma_b=3$ (black), $30$ (red), $80$ (blue), $300$ (green), and $500$ (violet). 
While $pp$ interactions dominate for $\Gamma_b=3$ and $30$, $p\gamma$ interactions
are responsible for shaping the neutrino spectrum in the case of $\Gamma_b=300$ and $500$.
The case with  $\Gamma_b=80$ is an intermediate case where both $pp$ and $p\gamma$ interactions
are effective, although only the latter component is
 visible here.
}
\end{figure}
Figure~\ref{fig:singlesource} shows the fluence of a burst at $z=1$ as a function of $\Gamma_b$ as from Eq.~(\ref{eq:fluence}) and after flavor oscillations.
For $\Gamma_b=3$ and 30 the spectrum is clearly dominated by $pp$
interactions, while for $\Gamma_b = 300$ and 500 by $p\gamma$ interactions.
The spectrum at $\Gamma_b=80$ is mainly determined by $pp$ interactions
at low energies (not visible in the plot because the flux is lower than
the bottom value of the y-axis of the plot) and by $p\gamma$ interactions in the region around
$10^7$~GeV.
The sharp rise of the neutrino spectrum at about $10^7$~GeV is due
to the fact that this object is optically thick ($\tau_T > 1$) and the
correspondent initial neutrino spectrum has a sharp rise due to the
black-body photon spectrum distribution (see Fig.~\ref{fig:nuinitial}).

%%%%%%%%%%%%%%%%%%%%%%%%%%%%%%%%%%%%%%%%%%%%%%%%%%%%%%%%%%%%%%%%%%%%%%
\section{Diffuse high-energy neutrino emission from astrophysical bursts}\label{sec:diffuse}
%%%%%%%%%%%%%%%%%%%%%%%%%%%%%%%%%%%%%%%%%%%%%%%%%%%%%%%%%%%%%%%%%%%%%%
\begin{figure}
\includegraphics[width=1.\columnwidth]{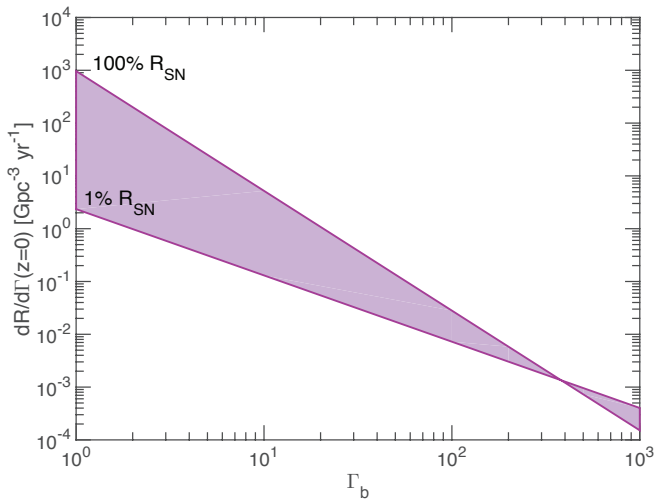}
\caption{\label{fig:localrate} Local formation rate of the  jets per unit volume per unit $\Gamma_b$, $R_j(z=0,\Gamma_b)$, as a function of $\Gamma_b$ for fixed $\rho_{0,{\rm HL-GRB}}$ and  $\zeta_{\rm SN} = 1, 100 \%$ respectively (see text for details). }
\end{figure}
We assume that the redshift evolution of baryon-rich and ordinary high-luminosity GRBs
is a function of the redshift and of the Lorentz boost factor;
$R_j(z,\Gamma_b)d\Gamma_b = R(z) \xi(\Gamma_b)d\Gamma_b$ is the
formation rate density of the bursts with the Lorentz factor between
$\Gamma_b$ and $\Gamma_b + d\Gamma_b$.
The redshift-dependent part of
$R_j(z,\Gamma_b)$ follows the star formation rate~\cite{Yuksel:2008cu}:
\begin{equation}
R(z)\propto \left[(1+z)^{p_1 k} + \left(\frac{1+z}{5000}\right)^{p_2 k}
	      + \left(\frac{1+z}{9}\right)^{p_3 k} \right]^{1/k}\,
\end{equation}
with $k = -10$, $p_1 = 3.4$, $p_2 = -0.3$, $p_3 = -3.5$,
 and is normalized such that $R(0) = 1$.
As for the $\Gamma_b$ dependence on the rate, we assume $\xi(\Gamma_b) = \Gamma_b^{\alpha_{\Gamma}} \beta_{\Gamma}$ and fix the parameters $\alpha_{\Gamma}$ and $\beta_{\Gamma}$ in such a way that 
\begin{eqnarray}
\int_1^{10^3} d\Gamma_b\ \Gamma_b^{\alpha_\Gamma} \beta_{\Gamma} &=& R_{\rm SN}(0) \zeta_{\rm SN} \frac{\theta_{\rm SN}^2}{2}\ , \\ 
\int_{200}^{10^3} d\Gamma_b\ \Gamma_b^{\alpha_\Gamma} \beta_{\Gamma} &=& \rho_{0,{\rm HL-GRB}}\ ,
\label{eq:jetrate}
\end{eqnarray}
 where $\zeta_{\rm SN}$ is the fraction of core-collapse SNe that develop  jets, $\theta_{\rm SN}^2/2$ the fraction of the jet pointing towards
us, $R_{\rm SN}(0) \simeq 2 \times
10^5$~Gpc$^{-3}$~yr$^{-1}$~\cite{Dahlen:2004km,Strolger:2015kra} the
local SN rate, and $\rho_{0,{\rm HL-GRB}} = 0.8$~Gpc$^{-3}$~yr$^{-1}$ being
 an optimistic estimation of the observed local
high-luminosity GRB rate~\cite{Wanderman:2009es}.
In order to give an idea of the dependence of $R_j(z,\Gamma_b)$ on $\Gamma_b$, Fig.~\ref{fig:localrate}
shows $R_j(z=0,\Gamma_b)$ as a function of $\Gamma_b$  for fixed $\rho_{0,{\rm HL-GRB}}$ and  $\zeta_{\rm SN} = 1, 100 \%$ respectively.

The total diffuse neutrino intensity from all bursts is therefore defined in the following way:
\begin{widetext}
\begin{eqnarray}
I(E_\nu) = \int_{\Gamma_{b,\rm min}}^{\Gamma_{b,\rm max}} d\Gamma_b \int_{z_{\rm min}}^{z_{\rm max}} dz \frac{c}{2 \pi \theta_j^2 H_0 \Gamma_b}
\frac{1}{\sqrt{\Omega_M (1+z)^3+\Omega_\Lambda}} R_j(z,\Gamma_b)  E^\prime_j f_p N_a [1-(1-\chi_p)^{\tau^\prime_p}] \left(\frac{dN_{\nu_\mu}}{dE^{\prime}_\nu}\right)_{\rm osc}\ . 
\end{eqnarray}
\end{widetext}

\begin{figure}[h]
\includegraphics[width=1.\linewidth]{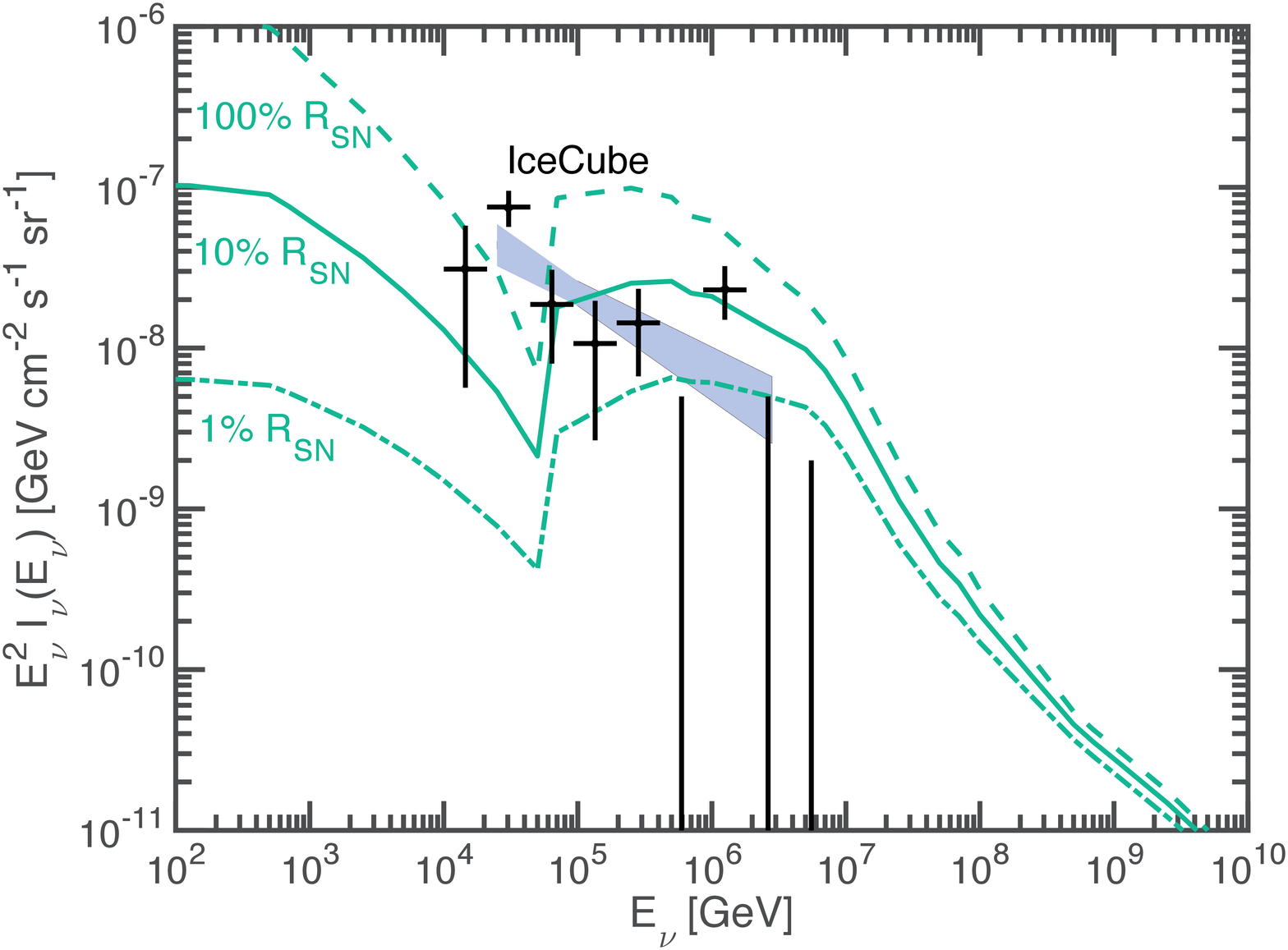}\\
\includegraphics[width=1.\linewidth]{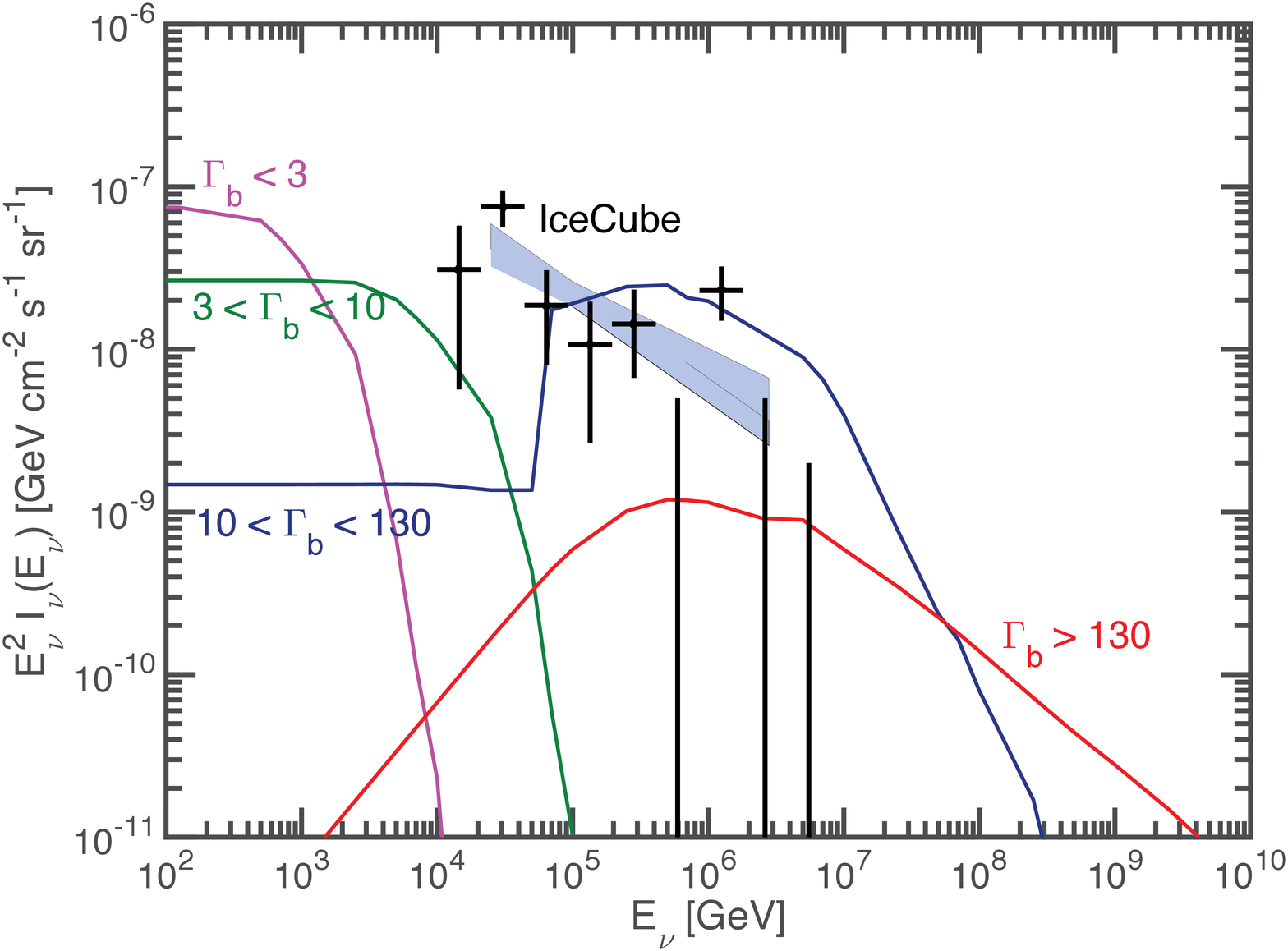}\\
\caption{\label{fig:diffuse}Top panel: Diffuse intensity for one
 neutrino flavor after flavor oscillations as a function of the energy and for  $\zeta_{\rm SN} =
 1$, 10, and 100\%, plotted with a dashed, solid and dot-dashed line,
 respectively. The blue band and the black data points correspond to the
best-fit power-law model and the IceCube data from Ref.~\cite{Aartsen:2015knd}. $\zeta_{\rm SN} = 100\%$ is
incompatible with the current IceCube data, while $\zeta_{\rm SN} = 10\%$
 is marginally allowed. Bottom panel: Partial contributions to the
 diffuse neutrino intensity for one neutrino flavor from different
 regimes of $\Gamma_b$, for $\zeta_{\rm SN} = 10\%$.
 As $\Gamma_b$ increases, the neutrino spectrum peaks at larger neutrino
 energies.}
\end{figure}

The top panel of Fig.~\ref{fig:diffuse} shows the total diffuse emission
from astrophysical bursts as a function of the neutrino energy for one
neutrino flavor obtained by assuming $[z_{\rm min},z_{\rm max}]=[0,7]$ and $[\Gamma_{b,\rm min},\Gamma_{b,\rm max}]=[1,10^3]$.
The continuous line stands for $\zeta_{\rm SN} = 10\%$, while
the dashed (dot-dashed) line is obtained by adopting  $\zeta_{\rm SN} = 100\% \ (1\%)$.
For comparison, the IceCube data as well as a band corresponding to the
single power-law fit~\cite{Aartsen:2015knd} are shown.
The figure shows that these jets could represent a major component of  the flux of
the IceCube neutrinos for $\zeta_{\rm SN} < 10\%$, especially in the PeV energy range.

Assuming that baryon-rich jets and ordinary GRBs belong all to the same family
and evolve by following $R_j(z,\Gamma_b)$, one can also indirectly
constrain the local rate of baryon-rich bursts by adopting the IceCube
high-energy neutrino data.
In fact, Fig.~\ref{fig:diffuse} suggests that a  local rate  of baryon-rich jets with 
$\zeta_{\rm SN}$ higher than tens of percent is excluded from the
current IceCube data set.  Our findings on the abundance of baryon-rich jets are also in agreement with the ones in Ref.~\cite{Kowalski:2014zda},
where the local abundance of transient sources of high-energy neutrinos is found to be lower than $10$~Gpc$^{-3}$ yr$^{-1}$ 
to do not contradict the non-observation of such sources in dedicated neutrino searches.

In order to disentangle the dependence of the neutrino diffuse intensity
from $\Gamma_b$, the bottom panel of  Fig.~\ref{fig:diffuse} shows
partial contributions to the total diffuse emission from different
regimes of $\Gamma_b$ for $\zeta_{\rm SN} = 10\%$.
As $\Gamma_b$ increases, the neutrino intensity peaks at higher
energies.
The flux for $\Gamma_b>130$ reproduces the expected  diffuse
intensity from high-luminosity GRBs in the PeV energy range; 
on the other hand, jets with $\Gamma_b < 10$ are responsible for a neutrino 
flux that is relevant in the TeV energy
range  (see also, e.g., Refs.~\cite{Tamborra:2015qza,Murase:2008sp} about the typical neutrino energy spectra from $p\gamma$ and $pp$ interactions).
For the assigned input parameters, astrophysical bursts with $10<\Gamma_b<130$ are responsible for a neutrino flux compatible with the current IceCube 
neutrino data set for particular values of $\zeta_{\rm SN}$.
Such jets belong to an intermediate class between choked and
high-luminosity GRBs, which is optically thick and in which $pp$ and
$p\gamma$ interactions are both relevant, as discussed in
Sec.~\ref{sec:neutrinoenergyspectrum}.

%%%%%%%%%%%%%%%%%%%%%%%%%%%%%%%%%%%%%%%%%%%%%%%%%%%%%%%%%%%%%%%%%%%%%%
\section{Uncertainties on the jet model parameters}\label{sec:radiation}
%%%%%%%%%%%%%%%%%%%%%%%%%%%%%%%%%%%%%%%%%%%%%%%%%%%%%%%%%%%%%%%%%%%%%%

The results presented in Sec.~\ref{sec:diffuse} have been obtained by assuming a simple model with common properties for all GRBs, except for the Lorentz factor $\Gamma_b$. Our conclusions are however limited by the astrophysical uncertainties.
For example, we assumed that the local rate of successful GRBs is given by  $\rho_{0,{\rm HL-GRB}} = 0.8$~Gpc$^{-3}$~yr$^{-1}$~\cite{Wanderman:2009es};
this is an optimistic assumption as the local rate could be as low as
$0.5$~Gpc$^{-3}$~yr$^{-1}$~\cite{Wanderman:2009es}. We also consider the
simplest possible scaling law of the local cosmic rate of astrophysical
jets as a function of $\Gamma_b$ (Eq.~\ref{eq:jetrate}), given the lack
of data; other possible scaling relations might describe better the real
GRB family. We currently do not have data to describe the engine behind
low-$\Gamma_b$ jets and extrapolate their properties from the ones
measured for successful jets. Future observations may help to reduce
such uncertainties~\cite{Meszaros:2015krr} that we currently expect
might be responsible for a variation of up to one order of magnitude or
two of the estimated best-fit value of the flux.

\begin{figure}
\includegraphics[width=1.\linewidth]{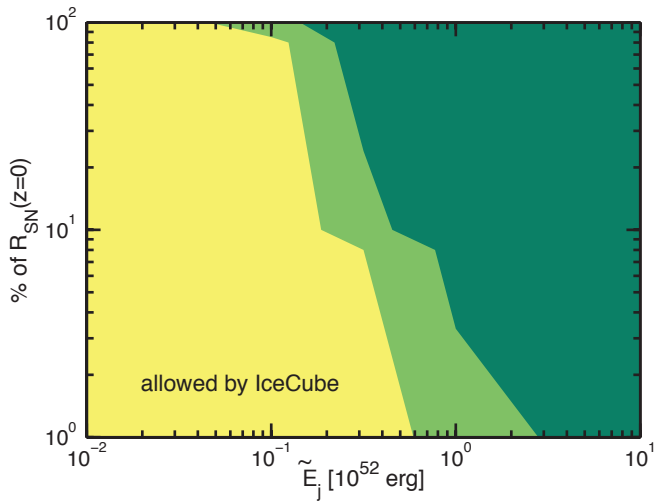}\\
\caption{\label{fig:contour}Contour plot of the allowed abundance of choked
 bursts expressed  as a 
 fraction of the local supernova rate that forms  choked jets, 
 $\zeta_{\rm SN}$, and as a  function of  the jet energy $\tilde{E}_j$. The yellow region is compatible with the IceCube data~\cite{Aartsen:2015knd}
 and the dark green one is excluded;  the light green region is marginally
 compatible.}
\end{figure}

Besides the local abundance of baryon-rich sources, the jet energy may be also a variable parameter.
Figure~\ref{fig:contour} represents $\zeta_{SN}$ as a function of
$\tilde{E}_j$. The contour plot shows the  allowed
 abundance of baryon-rich
bursts from the current IceCube high-energy neutrino data set~\cite{Aartsen:2015knd};  the yellow region is
compatible with the IceCube data, while the dark green one is excluded.\footnote{We define ``allowed region'' (``not-allowed region''), the region of the parameter space where 
$[E_\nu^2 I_\nu(E_\nu)]_{\rm theo} \le [E_\nu^2 I_\nu(E_\nu)]_{\rm IC, band}$ ($[E_\nu^2 I_\nu(E_\nu)]_{\rm theo} > [E_\nu^2 I_\nu(E_\nu)]_{\rm IC, band}$) for all energy points 
$E_\nu$ of the IceCube data;  
the ``marginally allowed region'' is the transition region of the parameter space
where roughly half of all energy points fall within one of the  two previous categories.}
A region of marginally allowed $(\tilde{E}_j,\zeta_{SN})$ falls in
between (plotted in light green).

Although the high-energy neutrino flux detected by the IceCube telescope is in the
same energy range where the neutrino emission from intermediate-$\Gamma_b$ jets
peaks, we are able to provide bounds on the local rate of baryon-rich GRBs
as a function of the jet energy by assuming a SN--GRB connection.
Such constraints are roughly comparable with the ones presented in
Ref.~\cite{Abbasi:2011ja}, obtained for choked sources. 
Note, however, that  the bounds on
$(\tilde{E}_j,\Gamma_b)$ in Ref.~\cite{Abbasi:2011ja} were extrapolated on the basis of  an
analysis on point sources, and  $\Gamma_b$ was considered as
fixed parameter typical of choked GRBs. Under the assumption of the SN-GRB connection, we 
expect that upper limits  on the abundance of choked sources are going to
become more stringent in the near future in the light of the increasing statistics 
of the IceCube data sets. 

We have assumed in this work that the parameters of each GRB are fixed during its duration. However, by including multiple internal shocks and assuming a non-narrow distribution 
in $\Gamma_b$  in each GRB as, e.g., in the simulations presented in Ref.~\cite{Bustamante:2014oka}, 
a qualitatively similar phenomenology to the one described above might be of relevance also in the transition from optically thick to optically thin emission regions. 

\begin{figure}[h]
\includegraphics[width=1.\linewidth]{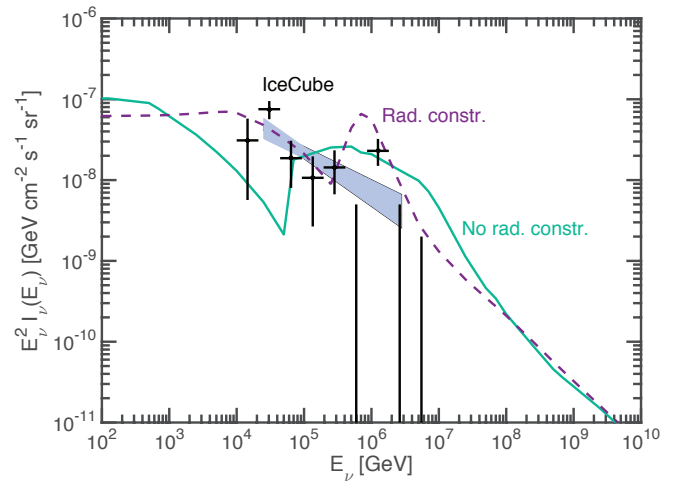}
\caption{\label{fig:radiation} Diffuse intensity for one
 neutrino flavor after flavor oscillations as a function of the energy and for  $\zeta_{\rm SN} =
 10$\%. The blue band and the black data points correspond to the
best-fit power-law model and the IceCube data from Ref.~\cite{Aartsen:2015knd}. 
 The green continue line represents the total neutrino intensity of our standard
 jet model, and the violet dashed line is the
 total neutrino intensity when radiation constraints are taken into account (see text for more details).}
\end{figure}

Thus far, we assumed that the burst duration is independent of  $\Gamma_b$. If high-energy particles are emitted mainly through internal shocks, the internal collisions occurring between the plasma shells inside the jet could however spread out in time and yield burst durations longer than the assumed $\tilde{t}_j=10$~s for bursts with low $\Gamma_b$. Although the dependence of $t_j$ on $\Gamma_b$ is relevant to describe the physics of astrophysical bursts, our assumption ($t_j(\Gamma_b)=$const.) does not affect our conclusions, see discussion about results presented in Fig.~\ref{fig:radiation}.

We work under the assumption that  internal collisionless shocks are able to accelerate protons efficiently for any Lorentz factor $\Gamma_b$. 
 As discussed in Refs.~\cite{Levinson:2007rj,Budnik:2010ru,Murase:2013ffa,Murase:2013hh}, this 
might not be the case if radiation-mediated shocks occur in choked sources; as a consequence, proton acceleration could not be as efficient as considered here and the
correspondent neutrino energy fluxes from baryon-rich sources might be affected. 
However,  as shown in Fig.~\ref{fig:diffuse}, the upper bound on $\zeta_{\rm SN}$   should not be affected, since it is only indirectly constrained from the IceCube data from jets with intermediate $\Gamma_b$ that
should, at least partially, evade the  radiation-dominated regime. In order to prove that, we include the condition to avoid radiation-mediated shocks for our
representative case with $\xi_{\rm SN}=0.1$, following the discussion in Ref.~\cite{Murase:2013ffa}. We vary the burst duration as a function of $\Gamma_b$ in
such a way to recover the conservative bound: $\tau_T \le 1$ (see
 Eq.~\ref{eq:tauT}) for any redshift $z$. Specifically, we consider
 $t_j=10$~s for $130 < \Gamma_b \le 10^3$, $t_j=500$~s for $50 <
 \Gamma_b \le 130$, and $t_j=10^6$~s for $\Gamma_b \le 50$. Note that
 such a choice of $t_j$ is also responsible for lower jet luminosities
 as $\Gamma_b$ decreases in agreement with the upper bounds on the
 luminosity shown in Fig.~3 of Ref.~\cite{Murase:2013ffa} in order to
 evade the radiation-dominated regime. The total neutrino intensity
 computed within this setup is plotted in Fig.~\ref{fig:radiation} (dashed
 violet curve) and it should be compared with the continuous green curve
 representing the total diffuse intensity for constant $t_j=10$~s also
 shown in the top panel of Fig.~\ref{fig:diffuse}. The condition $\tau_T
 \le 1$ affects the leading cooling processes discussed in
 Sec.~\ref{sec:proton} as a function of $\Gamma_b$ and the final shape
 of the expected neutrino intensity as shown in
 Fig.~\ref{fig:radiation}, but it does not drastically modify our
 conclusions.

Reference~\cite{Bromberg:2011fg} pointed out as the collimation shock in the jet inside the star might be of relevance for the  jet
dynamics and, as discussed in Ref.~\cite{Murase:2013ffa}, it might also affect the neutrino
production. On the basis of the study presented here, we conclude that our main results are robust even with respect to the inclusion of such effect now discarded. Nevertheless, further modeling of the jet properties by adding all the factors mentioned above is required, especially in light of the fact that jets with intermediate values of $\Gamma_b$ could produce a neutrino diffuse flux comparable to the one detected from IceCube.  

%%%%%%%%%%%%%%%%%%%%%%%%%%%%%%%%%%%%%%%%%%%%%%%%%%%%%%%%%%%%%%%%%%%%%%
\section{Conclusions}                \label{sec:conclusions}
%%%%%%%%%%%%%%%%%%%%%%%%%%%%%%%%%%%%%%%%%%%%%%%%%%%%%%%%%%%%%%%%%%%%%%
The most likely scenario explaining the formation of the long-duration
astrophysical bursts is the development of a jet out of a black hole or
an accretion disk, soon after the core collapse of a supernova.
However, observational evidence suggests that only a small fraction of  supernovae  evolves into high-luminosity gamma-ray bursts (GRBs)
with highly-relativistic jets.
Probably, softer jets, invisible or scarcely visible electromagnetically, could originate from the remaining optically 
thick supernova heirs. These objects are possibly even more abundant that
the ones leading to visible GRBs and  have been dubbed baryon-rich jets in this work. 

In this paper, we study the supernova--GRB connection, by assuming that ordinary high-luminosity GRBs and  baryon-rich jets originate from the same class of sources having
core-collapse supernovae as common progenitors. We hypothesize  that the local
rate of such sources decreases as the Lorentz boost factor $\Gamma_b$ increases. In order to investigate the neutrino emission 
from this class of astrophysical jets, we define a general neutrino emission model, including hadronuclear and photomeson interactions as well as cooling processes for mesons 
and protons. For simplicity, we assume that ordinary GRBs and baryon-rich jets have identical jet properties except for the Lorenz factor $\Gamma_b$, although we proof that our conclusions should not drastically change with respect to variations of the other model parameters.

We find that the neutrino fluence peaks in different energy ranges according to the Lorenz boost factor, ranging from  TeV energies for low-$\Gamma_b$ bursts
to PeV energies for high-$\Gamma_b$ bursts. The neutrino production in  low-$\Gamma_b$ jets is mainly due to hadro-nuclear interactions, while it is mainly determined by 
photon-meson interactions for bursts with high-$\Gamma_b$. 

The high-energy neutrino flux currently observed by the IceCube telescope could be generated, especially in the PeV region, from bursts with intermediate values of $\Gamma_b$
with respect to the typical ones of baryon-rich and bright GRBs: $\Gamma_b \in [10,130]$. Such sources with intermediate values of $\Gamma_b$ are optically thick, therefore
not or scarcely visible in photons, and $pp$ and $p\gamma$ interactions are both effective for what concerns the neutrino production.

Under the assumption that
supernovae evolve in astrophysical bursts with variable $\Gamma_b$, we point out that by comparing the  diffuse emission of high-energy neutrinos from jets with 
intermediate values of $\Gamma_b$ with the current best fit of the IceCube high-energy neutrino flux, one could put indirect constraints
on the local rate of choked GRBs. We find that  the present
IceCube data sets favor a local rate of choked sources lower than tens of
percent of the local core-collapse supernova rate. Such constraints are roughly
compatible with upper limits coming from dedicated searches on choked sources from the IceCube Collaboration. However,
we expect them to become tighter in the next future  in the light of the IceCube increasing statistics  and future generation neutrino telescopes~\cite{Aartsen:2014njl}. 

%%%%%%%%%%%%%%%%%%%%%%%%%%%%%%%%%%%%%%%%%%%%%%%%%%%%%%%%%%%%%%%%%%%%%%
\section*{Acknowledgments}
%%%%%%%%%%%%%%%%%%%%%%%%%%%%%%%%%%%%%%%%%%%%%%%%%%%%%%%%%%%%%%%%%%%%%%
We thank Imre Bartos, Mauricio Bustamante, Kohta Murase and Walter Winter for comments on the manuscript. This work was supported by the Netherlands Organization for Scientific
Research (NWO) through a Vidi grant.

%%%%%%%%%%%%%%%%%%%%%%%%%%%%%%%%%%%%%%%%%%%%%%%%%%%%%%%%%%%%%%%%%%%%%%
%%%  Bibliography  %%%%%%%%%%%%%%%%%%%%%%%%%%%%%%%%%%%%%%%%%%%%%%%%%%%
%%%%%%%%%%%%%%%%%%%%%%%%%%%%%%%%%%%%%%%%%%%%%%%%%%%%%%%%%%%%%%%%%%%%%%

%%%%%%%%%%%%%%%%%%%%%%%%%%%%%%%%%%%%%%%%%%%%%%%%%%%%%%%%%%%%%%%%%%%%%%
\end{document}